# Stochastic Model Predictive Control using Initial State Optimization

Henning Schlüter    Frank Allgöwer

*Institute for Systems Theory and Automatic Control, University of Stuttgart,
Pfaffenwaldring 9, 70550 Stuttgart, Germany
(e-mail: {schlueter, allgoewer}@ist.uni-stuttgart.de).*

**Abstract:** We propose a stochastic MPC scheme using an optimization over the initial state for the predicted trajectory. Considering linear discrete-time systems under unbounded additive stochastic disturbances subject to chance constraints, we use constraint tightening based on probabilistic reachable sets to design the MPC. The scheme avoids the infeasibility issues arising from unbounded disturbances by including the initial state as a decision variable. We show that the stabilizing control scheme can guarantee constraint satisfaction in closed loop, assuming unimodal disturbances. In addition to illustrating these guarantees, the numerical example indicates further advantages of optimizing over the initial state for the transient behavior.

## 1. INTRODUCTION

Model predictive control (MPC) represents an effective control technique for reliably handling constraints, guaranteeing their satisfaction even in the presence of disturbances and uncertainty. For nominal MPC schemes without disturbances and robust MPC schemes considering worst-case bounds on disturbances, rigorous theoretical guarantees, such as recursive feasibility, stability, and constraint satisfaction, are well established in the literature (Rawlings et al., 2017) yielding widely applied control schemes (Raković and Levine, 2019).

Prominent results in robust MPC are by Mayne et al. (2005) and Chisci et al. (2001), which propose efficient methods for linear systems subject to bounded disturbances based on constraint tightening. One key difference between the two is the choice of how to select the initial state for prediction. Where Chisci et al. (2001) uses the measured state directly, Mayne et al. (2005) instead treats the initial state also as a decision variable of the optimal control problem constrained with respect to the measurement. In robust MPC today, even very general classes of nonlinear systems can be considered using either choice (Raković and Levine, 2019).

However, many problems have more detailed information about the disturbance available, commonly described by its stochastic properties. For these systems, stochastic MPC (SMPC) schemes can reduce conservatism by taking knowledge about the distribution into account (Kouvaritakis and Cannon, 2016). Mesbah (2016) classifies these methods into two main categories. On one hand, there are stochastic programming-based approaches, commonly using randomized methods relying on the generation of many scenarios; and on the other hand, there are tube-based approaches, for example by Cannon et al. (2011), Hewing and Zeilinger (2018) and the scheme we will present. These approaches use analytic approximations to tighten the constraints of a nominal MPC, which is then used as a surrogate for the original problem to be solved.

The commonality between these methods is that they are based on the formulation by Chisci et al. (2001), which uses the measured state directly as the initial state. This formulation, however, commonly limits SMPC schemes to consider only bounded disturbances, even when only considering probabilistic constraints (Cannon et al., 2011; Lorenzen et al., 2017), since unbounded disturbances inevitably lead to infeasibility of the deterministic surrogate MPC. On the contrary, the most common stochastic distribution, namely, the Gaussian distribution, has infinite support and thus, cannot be used without truncation in these methods.

Some methods have been proposed to address these shortcomings. Farina et al. (2013) proposes to use a fallback strategy in case the MPC becomes infeasible with the measured state as the initial state for prediction, relying on the previously predicted state instead. Using probabilistic reachable sets (PRS) for the tightening, Hewing and Zeilinger (2018) show that the constraints are satisfied in closed loop despite also relying on a fallback strategy, which is a concept we will reuse for our own proof thereof. Instead of merely selecting between two possible initial states based on feasibility, we propose to optimize over the initial state similar to the robust MPC approach by Mayne et al. (2005).

Taking the fallback idea a step further in a different direction, Hewing et al. (2020) propose to always use previously predicted state as the new initial state as far as the constraints are concerned. A different recent approach by Yan et al. (2018) guarantees feasibility despite unbounded disturbances for suitably discounted violation probabilities. Both of these come with significant change to the interpretation of the chance constraints, either by considering the open-loop prediction, i.e., conditioning on the initial state of the system, or by discounting the violation. While differing in nature, both de-emphasize violations in the far future. To a lesser extent, this will also be the result of our scheme, as using an initial MPC state different from the measured state necessarily requires a relaxation of the condition of the constraints. While we will choose the measured state less frequently than the fallback scheme, the chosen state will usually be closer to the measurement than the open-loop prediction. Recently, a related initial state interpolation has also been independently proposed in Köhler and Zeilinger (2022).

★ Received on Feb. 17, 2022; Accepted June 6, 2022; Revised July 18, 2022. The authors thank the German Research Foundation (DFG) for financial support under the Grant GRK 2198/1-277536708, and Grant AL 316/12-2-279734922; and the International Max Planck Research School for Intelligent Systems (IMPRS-IS) for supporting Henning Schlüter.

Our main contribution is using the initial state as a decision variable of the optimal control problem for stochastic MPC while maintaining the common guarantees. We will show that this approach avoids the infeasibility issues with unbounded disturbances, is quadratically stabilizing when using the expected LQR cost, and guarantees constraint satisfaction in closed-loop for linear systems affected by general unimodal additive disturbances.

The problem setup is formulated in Sec. 2 with some implementation details for the underlining assumption deferred to Sec. 5. In Sec. 3, we present the MPC scheme and discuss our design choices for initializing the state prediction. The resulting guarantees are proven in Sec. 4 and illustrated with a numerical example in Sec. 6, which also reveals an improved transient behavior when compared to the approach by Hewing and Zeilinger (2018).

*Notation:* We use $x_i(k)$ to denote $x$ predicted $i$ time steps ahead from time $k$ and $x(k) = x_0(k)$ for quantities realized in close-loop. The probability of $A$, the expected value of $x$ and the variance of $x$ are written as $\mathbb{P}(A)$, $\mathbb{E}[x]$, and $\mathbb{V}[x]$, respectively. The weighted 2-norm is denoted as $\|x\|_Q^2 = x^\top Q x$ for positive definite matrices $Q$ and a sequence as $x_{0:N} := \{x_0, \ldots, x_N\}$. The Pontryagin set difference is $A \ominus B = \{a \mid a + b \in A \ \forall b \in B\}$.

## 2. PROBLEM SETUP

We address the regulation problem of a linear time-invariant system under additive stochastic disturbances

$$x(k+1) = Ax(k) + Bu(k) + w(k) \quad (1)$$

with state $x(k) \in \mathbb{R}^{n_x}$, input $u(k) \in \mathbb{R}^{n_u}$, and disturbances $w(k) \in \mathbb{R}^{n_x}$. The disturbances are independent and identically distributed with distribution $w(k) \sim \mathcal{Q}^w$, which can have infinite support.

The system is subject to separate chance constraints on the state and the input, i.e.,

$$\mathbb{P}(x(k) \in \mathcal{X}) \geq p_x \quad (2a)$$
$$\mathbb{P}(u(k) \in \mathcal{U}) \geq p_u \quad (2b)$$

where $\mathcal{X}$ and $\mathcal{U}$ are convex sets containing the origin. These probabilities are conditional given some past state, which is specified later in Sec. 4.1. While (2) includes the case of hard constraints by imposing the probability of 1, these can usually only be satisfied for bounded disturbances.

The recursive feasibility of the proposed algorithm will require the disturbance distribution to be multivariate unimodal, in particular we use central convex unimodality as defined by (Dharmadhikari and Jogdeo, 1976, Def. 3.2).

**Definition 1.** (Central convex unimodality). A distribution $\mathcal{Q}$ in $\mathbb{R}^n$ is called central convex unimodal if it is in the closed convex hull of the set of all uniform distributions on symmetric compact convex bodies in $\mathbb{R}^n$.

This is a fairly weak assumption, satisfied by many relevant distributions. For example, all log-concave distributions, i.e., distributions having a probability density function $\exp(\phi)$ where $\phi$ is a concave function, are central convex unimodal (Saumard and Wellner, 2014). In particular, the multivariate Gaussian distribution is central convex unimodal.

As in (Kouvaritakis and Cannon, 2016), the controller objective is to minimize the expected infinite-horizon LQR cost

$$J = \mathbb{E}\left[\sum_{i=0}^{\infty} x(k)^\top Q x(k) + u(k)^\top R u(k) - \ell_{ss}\right], \quad (3)$$

where the expected optimal steady-state stage cost $\ell_{ss}$ is subtracted in each stage to ensure that this sum is finite.

### 2.1 Probabilistic reachable sets

As is common for tube-based SMPC (Mesbah, 2016), our algorithm uses an LQR controller $K$ as prestabilizing controller to design the constraint tightening, since this is the optimal controller for the unconstrained optimal control problem for (3).

Similar to (Hewing and Zeilinger, 2018), we use a constraint tightening based on PRS. Thus, we introduce a set $\mathcal{R}$, which satisfies for all $n \geq 0$

$$x(0) = 0 \implies \mathbb{P}(x(n) \in \mathcal{R}) \geq p \quad (4)$$

for the prestabilized system

$$x(k+1) = A_K x(k) + w(k). \quad (5)$$

Then, this set is called a PRS of probability level $p$. More details regarding the implementation of this assumption are discussed later in Sec. 5.

## 3. STOCHASTIC MPC

Using this constraint tightening we now present a stochastic MPC approach using a dual mode prediction scheme for the infinite-horizon cost as in (Kouvaritakis and Cannon, 2016). When considering constraints for an uncertain system, it is convenient to decompose the dynamics as:

$$z_{k+1} = Az_k + Bv_k \quad (6)$$
$$e_{k+1} = A_K e_k + w_k \quad (7)$$

splitting the state $x = z + e$ into a nominal and an error part. Since the initial error can be chosen as a stationary process starting at zero, its dynamics can be fully computed offline and captured for enforcing the chance constraints (2) with the help of a PRS.

Therefore, we can design a nominal MPC for $z$, while using the auxiliary LQR controller $K$ to keep the error $e$ small. Hence, we use the input signal

$$u(k) = v_0(k) + Ke(k), \quad (8)$$

where $v_0(k)$ is the first input computed by the nominal MPC.

### 3.1 Constraint tightening

With the decomposition (8) of the input, we can reformulate the chance constraints on $x$ and $u$ into deterministic constraints on $z$ and $v$ by substituting it into (2a), which yields

$$p_x \leq \mathbb{P}(z_i + \mathcal{R}_x \subseteq \mathcal{X} \wedge e_i \in \mathcal{R}_x) \qquad (9)$$
$$\leq \mathbb{P}(z_i + e_i \in \mathcal{X}) = \mathbb{P}(x_i \in \mathcal{X}),$$

i.e., a deterministic constraint for the nominal state $z_i$ with a PRS $\mathcal{R}_x$ of probability $p_x$ for the error $e_i$. Proceeding analogously for the input with a PRS $\mathcal{R}_u$ of probability $p_u$, yields the deterministic surrogate constraints

$$z_i \in \mathcal{Z} \coloneqq \mathcal{X} \ominus \mathcal{R}_x, \qquad (10a)$$
$$v_i \in \mathcal{V} \coloneqq \mathcal{U} \ominus \mathcal{R}_u. \qquad (10b)$$

Similarly, different constraints can be included by introducing additional PRS, for example, individual chance constraints opposed to the joint constraints considered here.

### 3.2 Initialization

A fundamental issue with having constraints in case of unbounded disturbances, is that any disturbance might take the state anywhere in $\mathbb{R}^{n_x}$. In such an event, the optimal control problem starting at the new state may become infeasible. Yet, we still need to be able to compute an input, as the controller should not fail for disturbances already considered in the model.

While particularly problematic when using deterministic surrogates, these feasibility issues are already inherent in the chance constraints. If these are, for example, conditioned on the previously measured state, as is common for SMPC with bounded disturbances, that state has to lie inside some strict subset of $\mathbb{R}^{n_x}$ such that the constraints are feasible. For bounded disturbances, this can be taken into account with the tightening, such that the problem remains feasible even when conditioned on a new state, since the new state is in a bounded neighborhood of the prediction.

For unbounded disturbances, however, this is not possible, as the disturbed state cannot be bounded against the prediction. As such, alternatives have emerged for unbounded disturbances. For example, Farina et al. (2013) and Hewing and Zeilinger (2018) use a conditional update rule for the initial state $z_0$ used in the nominal MPC, which uses the measurement $x(k)$ whenever feasible, while otherwise relying on the predicted state $z_1(k-1)$. Then, by the choice of an appropriate terminal constraint, recursive feasibility becomes trivial. While convenient for deriving guarantees, the state from the last prediction may not be the optimal initial state for the nominal MPC, if the measured state is infeasible.

Therefore, we propose to optimize over the initial condition of the MPC, similar to the idea from Mayne et al. (2005) for robust MPC, albeit with a slightly different motivation. While there are several common options for constraining the initial state in the optimization, the obvious choices fail.

Firstly, motivated by robust MPC approaches, one could consider $x(k) - z_0(k) \in \mathcal{R}$ with the set from the constraint tightening. This, however, does not work as the PRS does not contain all possible disturbances. Thus, the optimization will still become infeasible for large enough disturbances.

Secondly, using a set that contains all possible disturbances is also meaningless, since in general with unbound distribution, this set will be the entire $\mathbb{R}^{n_x}$, leaving the initial state unconstrained. Hence, the trivial zero solution would always be optimal, making the MPC superfluous.

Instead, we constrain the initial state to lie on the straight line connecting the closed-loop state $x(k)$ and the prediction $z_1(k-1)$ for it, which translates into the constraint

$$\exists\, 0 \leq \xi \leq 1 : z_0(k) = (1-\xi)x(k) + \xi z_1(k-1). \qquad (11)$$

It is crucial that this set contains the previous prediction for the state $z_1(k-1)$, as this will enable us to derive guarantees using the previous solution as candidate solution, similarly to the work of Farina et al. (2013). Including the closed-loop state $x(k)$ is beneficial since this is the actual state to be regulated. Further, using $z_0(k) = x(k)$ allows us to reset the condition of the probabilities to the current time as the error term becomes zero. Besides being the minimal connected set the choice of a line is key in deriving the guarantees.

### 3.3 Stochastic MPC Formulation

Bringing the ideas from the previous sections together, the deterministic surrogate optimal control problem for the nominal system is

$$\min_{v(k),z(k),\xi(k)} J(x(k), [v - Kz]_{0:N-1}(k)) + \ell_\xi(\xi(k)) \qquad (12a)$$
$$\text{s.t.} \quad z_{i+1}(k) = Az_i(k) + Bv_i(k), \qquad (12b)$$
$$z_i(k) \in \mathcal{Z}, \qquad (12c)$$
$$v_i(k) \in \mathcal{V}, \qquad (12d)$$
$$z_0(k) = (1-\xi(k))x(k) + \xi(k)z_1(k-1), \qquad (12e)$$
$$z_N(k) \in \mathcal{Z}_F, \qquad (12f)$$
$$v_j(k) = Kz_j(k), \forall j \geq N \qquad (12g)$$
$$\xi(k) \in [0,1], \qquad (12h)$$
$$\forall i \in \{0,...,N-1\} \qquad (12i)$$

with state and input sequence $z_{0:N}$, $v_{0:N-1}$, using the tightened constraint from (10), and the initial condition from (11). The terminal set $\mathcal{Z}_F$ is subject to the usual requirements that it satisfies $(A+BK)\mathcal{Z}_F \subseteq \mathcal{Z}_F \subseteq \mathcal{Z}$ and $K\mathcal{Z}_F \subseteq \mathcal{V}$, i.e., is positive invariant under the control and fulfills the constraints. The cost function $J(x, c_{0:N-1})$ will be introduced in Lemma 6 as an analytic expression for the expected infinite-horizon cost (3) taking the auxiliary controller $K$ for the error into account. Thus, it can also serve as terminal cost as is common for dual mode prediction schemes. The optional cost term $\ell_\xi$ additionally penalizes the use of an initial nominal state different from the measured state, which may be of interest in light of the conditioning in Cor. 5. However, the term $\ell_\xi$ is neither required for nor jeopardizing the guarantees.

### 4. THEORETICAL GUARANTEES

The foundation of the theoretical results developed in this section is that (12) is recursively feasible, which is immediate by construction.

**Lemma 2.** If problem (12) is feasible for $k=0$ with $z_1(-1) = x(0)$, then it is recursively feasible.

*Proof.* The shifted previous solution $v_i(k) = v_{i+1}(k-1)$ will be feasible with $\xi(k) = 0$ by the requirements on $\mathcal{Z}_F$. $\square$

This specific initialization leads to $z_0 = x(0)$. In principle, other values for $z_1(-1)$ could be used to achieve initial feasibility. However, by using $z_0 = x(0)$ we have $\xi(0) = 0$ which is the relevant for the conditioning of the chance constraints as we will see in the next section, in particular for Cor. 5.

## 4.1 Chance Constraint Satisfaction in close-loop

The constraint tightening (10) by itself only yields chance constraint satisfaction in prediction when $\xi = 0$. However, closed-loop chance constraint satisfaction requires to consider all $\xi \in [0, 1]$. For that we require the following result on convolution of distributions.

**Lemma 3.** Let the random variables $w$ and $x$ be independent and central convex unimodal, then the probability $\mathbb{P}(w + kx \in \mathcal{R})$ is non-increasing in $k \in [0, \infty]$, for any convex symmetric set $\mathcal{R}$.

*Proof.* The result follows immediately from Thm. 3.3 and Thm. 3.4 by Dharmadhikari and Jogdeo (1976). □

This lemma allows us to show that $\mathcal{R}$ is a PRS for the closed-loop error $e(k)$, which implies chance constraint satisfaction for the closed loop, similar to (Hewing and Zeilinger, 2018).

**Theorem 4.** Let $Q^w$ be central convex unimodal and let $\mathcal{R}$ be a convex symmetric set. For system (1) under the control law (8) resulting from (12) with tightening (10), we have

$$\mathbb{P}(e(k) \in \mathcal{R}) \geq \mathbb{P}(e_k(k_0) \in \mathcal{R}) \quad (13)$$

conditioned on $e(k_0) = 0$, for all $k \geq k_0$.

*Proof.* The predicted error $e_i(k)$ depends on the closed-loop disturbances $w(k_0), \ldots, w(k-1)$ and the future disturbances $w_0(k), \ldots, w_{i-1}(k)$ via prediction. To separate these influences split the error $e_n(k) = A_K^n e(k) + \tilde{e}_n(k)$, by introducing the term $\tilde{e}_n(k) = \sum_{i=k_0}^{N-1} A_K^{n-i-1} w_i(k)$, $\tilde{e}_0(k) = 0$. Now, we show that

$$\mathbb{P}(e_n(k-n) \in \mathcal{R}) \geq \mathbb{P}(e_{n+1}(k-n-1) \in \mathcal{R})$$

for $n = 0, \ldots, k-1$, which applied recursively yields the statement. Thus, by substituting the error we obtain

$$\mathbb{P}(e_n(k-n) \in \mathcal{R}) = \mathbb{P}\big(A_K^n e(k-n) + \tilde{e}_n(k-n) \in \mathcal{R}\big)$$
$$= \mathbb{P}\big(\xi A_K^{n+1} e(k-n-1) + \xi A_K^n w(k-n-1) + \tilde{e}_n(k-n) \in \mathcal{R}\big)$$

with $e(k-n) = \xi A_K e(k-n-1) + \xi w(k-n-1)$. Then since $\tilde{e}_n$, $e(k-n-1), w(k-n-1)$ are convex unimodal and independent, Lemma 3 allows to lower-bound this by

$$\geq \mathbb{P}\big(A_K^{n+1} e(k-n-1) + A_K^n w(k-n-1) + \tilde{e}_n(k-n) \in \mathcal{R}\big).$$

As $\tilde{e}_{n+1}(k-n-1)$ and $A_K^n w(k-n-1) + \tilde{e}_n(k-n)$ are equally distributed, this then equals

$$= \mathbb{P}\big(A_K^{n+1} e(k-n-1) + \tilde{e}_{n+1}(k-n-1) \in \mathcal{R}\big)$$
$$= \mathbb{P}(e_{n+1}(k-n-1) \in \mathcal{R}). \quad \square$$

**Corollary 5.** Theorem 4 implies satisfaction of (2) conditioned on any $x(k_0)$, where $\xi(k_0) = 0$, for the closed-loop system.

*Proof.* Since for all $k \geq k_0$ Thm. 4 implies $\mathbb{P}(e(k) \in \mathcal{R}_x) \geq p_x$ and $\mathbb{P}(Ke(k) \in K\mathcal{R}_u) \geq p_u$ and Lemma 2 yields $z(k) \in \mathcal{X} \ominus \mathcal{R}_x$ and $v(k) \in \mathcal{U} \ominus K\mathcal{R}_u$, the claim follows as in (9). □

Thus, using an initialization as in Lemma 2 with $\xi(0) = 0$ will guarantee constraint satisfaction conditioned on initial time in the same sense as indirect feedback SMPC by Hewing et al. (2020), which lacks feedback on the constraints. Our approach, however, takes measurements as far as possible into account for the constraints. Thus, the scheme is not only able to benefit from any disturbance taking the state away from the constraint, but also ensures constraint satisfaction conditioned on the last time, where $\xi = 0$ was chosen.

## 4.2 Stability and Cost decrease

Using the same cost function (3) as (Kouvaritakis and Cannon, 2016, Alg. 7.1), allows us to reuse their stability proof without significant change. Firstly, an analytic solution for (3) under (8) follows immediately from (Kouvaritakis and Cannon, 2016, Thm. 6.1 & Cor. 6.1).

**Lemma 6.** (Predicted cost). Let $P$ be the solution to the Riccati equation corresponding to the LQR controller $K$. Then the system (1) under the control law (8) using $c_i = v_i - Kz_i$ has the predicted cost (3) given by

$$J(x(k), c_{0:N}(k)) = \|x(k)\|_P^2 - \mathrm{tr}\,\Sigma_\infty P + \sum_{i=0}^{N-1} \|c_i(k)\|_{R+BPB}^2 \quad (14)$$

with $\Sigma_\infty$ as solution of $\Sigma_\infty = (A + BK)\Sigma_\infty (A + BK)^\top + \mathbb{V}[w]$.

The first two terms are of course constant w.r.t. the optimization variables in (12), thus only the last term

$$\sum_{i=0}^N \|v_i(k) - Kz_i(k)\|_{R+BPB}^2 \quad (15)$$

has to be included there. However, for ease of interpretation we consider the full cost. Then, use of the full infinite-horizon cost combined with Lemma 2 allows proofing closed-loop stability, with $\mathbb{E}_k$ denoting the expected value conditioned on time $k$.

**Theorem 7.** The closed-loop system (1) under the control law (8) resulting from (12) satisfies the quadratic stability condition

$$\lim_{r \to \infty} \frac{1}{r} \sum_{k=0}^r \mathbb{E}_0\big[\|x_k\|_Q^2 + \|u_k\|_R^2\big] \leq \ell_{ss}. \quad (16)$$

*Proof.* By optimality and using (3), we have
$$\mathbb{E}_k[J^*(x(k+1))] \leq \mathbb{E}_k[J(x(k+1), c_{k+1:k+N+1}(k))]$$
$$\leq J^*(x(k)) - (\|x(k)\|_Q^2 + \|u(k)\|_R^2 - \ell_{ss}),$$

where $J^*$ denotes the cost of the optimal solution. Since $J^*$ is finite, taking the expectation conditioned on $x_0$ yields the result, for details see (Kouvaritakis and Cannon, 2016, Thm. 7.1). □

Having proven the required guarantees for the MPC scheme, i.e., recursive feasibility in Lemma 2, constraint satisfaction in closed loop with Cor. 5, and stability by Thm. 7, we will illustrate the properties with numerical example in Sec. 6, but before that, there are some implementation details regarding PRS to discuss in the next section.

## 5. IMPLEMENTATION

In this section, will discuss how to compute the symmetric PRS $\mathcal{R}_x$, $\mathcal{R}_u$ and the terminal set $\mathcal{Z}_F$.

To compute probabilistic reachable sets (4), we first require the notion of probabilistic $n$-step reachable sets, which satisfy

$$x(0) = 0 \implies \mathbb{P}(x(n) \in \mathcal{R}^n) \geq p \quad \text{(see 4)}$$

for one particular $n \geq 0$. Thus, the union of all n-step PRS defines a PRS $\mathcal{R} \supseteq \bigcup_{n=0}^\infty \mathcal{R}^n$. Hewing and Zeilinger (2018, Lemma 2) have shown that symmetric $n$-step PRS are nested for central convex unimodal disturbances, i.e., every symmetric $n$-step PRS is also an $n-1$-step PRS. Thus, it is sufficient for finding a PRS to compute an $n$-step PRS for

$n \to \infty$. In this work, we only require the symmetric case where $\mathbb{E}[x] = 0$, as the error as consider in Thm. 4 always has expected value zero. Hence, we restrict our discussion to that case.

Then, the steady-state variance $\Sigma_\infty = \lim_{k\to\infty} \mathbb{V}[x(k)]$ of the prestabilized system (5) can be computed from

$$\mathbb{V}[x(k+1)] = A_K \mathbb{V}[x(k)] A_K^\top + \mathbb{V}[w(k)] \quad (17)$$

or in the special case of a normally distributed disturbances the steady-state variance $\Sigma_\infty \succ 0$ is the solution to the Lyapunov equation

$$0 = A_K \Sigma_\infty A_K^\top - \Sigma_\infty + \mathbb{V}[w(k)]. \quad (18)$$

Using Chebyshev's inequality (Chebyshev, 1867) we obtain an ellipsoidal PRS of probability level $p$

$$\mathcal{R}_x := \{x \mid x^\top \Sigma_\infty^{-1} x < \tilde{p}\} \quad (19)$$

with $\tilde{p} = \frac{n_x}{1-p}$. For Gaussian disturbances this bound can be improved to $\tilde{p} = \chi_{n_x}^2(p)$ using the inverse cumulative distribution function of the chi-squared distribution with $n_x$ degrees of freedom.

While the set $K\mathcal{R}_u$ (cf. (10b)) can be simply computed as $\{Kx \mid x \in \mathcal{R}_u\}$ using the same method as for the state, it will be significantly less conservative to first compute the distribution of the input. Then, the PRS can directly be constructed with variance $\mathbb{V}[u] = K\mathbb{V}[x]K^\top$.

Then, the terminal set $\mathcal{Z}_F$ can be computed based on the tightened constraint sets using standard methods, for example using (Kouvaritakis and Cannon, 2016, Alg. 2.1).

## 6. NUMERICAL SIMULATION

A widely used benchmark case study in the SMPC literature is the DC-DC-converter regulation problem. The corresponding linear dynamics of the form (1) with

$$A = \begin{bmatrix} 1 & 0.0075 \\ -0.143 & 0.996 \end{bmatrix}, \quad B = \begin{bmatrix} 4.798 \\ 0.115 \end{bmatrix}, \quad (20)$$

as previously considered by Cannon et al. (2011); Lorenzen et al. (2017). The SMPC cost weights are $Q = \text{diag}[1, 10]$ and $R = 10$ and the prediction horizon is $N = 8$. Since the algorithm can handle unbounded disturbances, we deviate from previous works and assume a Gaussian distribution without truncation for the disturbances. To better distinguish between the algorithms, we also use an increased variance of $\Sigma_w = 0.1\, \mathbb{I}_2$. To study the constraint satisfaction we consider a single chance constraint on the first state

$$\mathbb{P}(x^1(k) \leq 2) \geq 0.6 \quad (21)$$

for the system with an initial state $x(0) = \begin{bmatrix} 0.6455, & 1.3751 \end{bmatrix}^\top$ at the border of the initially feasible region, cf. Lemma 2.

Then using (19), we obtain the tightened constraint as

$$z^1 \leq 0.6455 \quad (22)$$

and a terminal set

$$\mathcal{Z}_F := \left\{ z \, \middle| \, \begin{bmatrix} 1.0000 & 0.0000 \\ -0.1559 & 1.8933 \end{bmatrix} z \leq 0.6455 \cdot \mathbb{1}_2 \right\}. \quad (23)$$

We compare our scheme with a variant of (Hewing and Zeilinger, 2018) using the same cost function as our scheme. Then, the main difference lies in the choice of initial condition

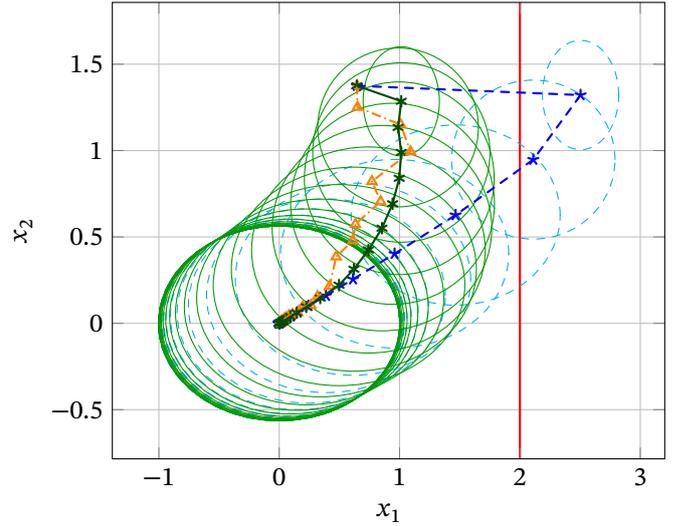

Figure 1. Plot of averaged closed-loop response of SMPC-ic in green (—∗—) compared to LQR as dashed blue lines (- -∗- -) with ellipses indicating one standard deviation. For SMPC-bak (-·-▲-·-) only the mean response is shown.

of the MPC (12), where Hewing and Zeilinger (2018) use a backup strategy

$$z_0(k) = \begin{cases} x(k), & \text{if feasible in (12)}, \\ z_1(k-1), & \text{otherwise}, \end{cases} \quad (24)$$

and thus do not require the optimization variable $\xi$. We will call this scheme using (24) SMPC-bak and our scheme using (11) SMPC-ic. For both we use $\ell_\xi \equiv 0$. However, as discussed in Sec. 4.1 for considering constraint satisfaction an additional cost term $\ell_\xi$ might be beneficial. Thus, we additionally consider SMPC-$\ell_\xi$ as a variant of our SMPC-ic with a linear cost $\ell_\xi(\xi) = 1600\xi$ adding a strong preference for choosing the feasible initial state closest to the measured state.

*Computation times:* The implementation for this simulation uses PICOS (Sagnol and Stahlberg, 2022) and Mosek (MOSEK ApS, 2022) under Python 3.10. Using an Intel® Core™ i9-10900 CPU at 2.80 GHz we obtain an average time across 250 000 runs of 6.7 ± 2.7 ms for SMPC-ic, while SMPC-bak requires 9.3 ± 4.6 ms, which is a 39 % increase on average and 57 % in standard deviation. The time saving can be easily attributed to the fact that SMPC-ic only needs to solve the optimization problem (12) once whereas SMPC-bak might need to solve (12) twice. This benefit is especially clear when the system is close to the constraint, when SMPC-bak relies on the backup strategy. When operated far way from the constraint, one might expect that SMPC-bak is slightly quicker since its optimization problem has one less variable. However, these savings were negligible in our simulations only amounting to 0.2 ± 4 ms.

*Constraint satisfaction in closed-loop:* Given the initial feasibility requirement, the initial state has to satisfy the tightened constraint. Hence, to still illustrate constraint satisfaction, we rely on the dynamics to violate the constraint in one step under unconstrained optimal control. This leads a significant constraint violation of 95 %, i.e., $\mathbb{P}_0(x^1 \leq 2) < 0.05$, for the LQR controller. Therefore, the chance constraint will be active for the SMPC schemes.

Table 1. Results of Monte Carlo simulation of (20).

|  | constraint violation $\mathbb{P}_0(x^1 > 2)$ | relative cost $\|x(0)\|_{\bar{P}}^{-2} J$ |
| --- | --- | --- |
| LQR | 94.47 % | 100.00 % |
| SMPC-ic | 10.79 % | 109.76 % |
| SMPC-$\ell_\xi$ | 9.19 % | 113.95 % |
| SMPC-bak | 8.39 % | 112.95 % |

In Fig. 1, we contrast SMPC-ic with the LQR trajectories as obtained from a Monte Carlo simulation over 250 000 different disturbance realizations. Firstly, we observe that clear violation of constraints is avoided by the SMPC-ic. As expected, our SMPC-ic slides along the tightened constraint for the first four steps to be as close to optimal as possible. While the closed-loop state in Fig. 1 seems to hover around 1 in the figure, the nominal state observes the 0.6455 of the tightened constraint as expected. The SMPC-bak fails to stick to the constraint, as it has to flip-flop between its two modes. With $x(k)$ not being consistently feasible at the constraint and $z_1(k-1)$ often being quite different, this causes the undesirable jitter observed in Fig. 1.

Secondly, we observe that the constraint tightening is quite conservative for the example. Given the full results of the Monte Carlo simulation in Tab. 1, we see that this is inherent with fixed size tubes based on PRS for this example. Overall the relation between conservatism of the variants of SMPC considered is representative also for other examples. That is SMPC-ic is less conservative than either SMPC-$\ell_\xi$ or SMPC-bak, which usually preform quite similar with SMPC-$\ell_\xi$ being only marginally less conservative. There are other examples, as the one used by Hewing and Zeilinger (2018), where all schemes are less conservative.

*Closed-loop performance:* Returning to Fig. 1, we can observe that the SMPC-ic is stabilizing and matches the LQR asymptotically, as expected by Thm. 7. In Tab. 1, we see that similar to the RMPC case (Mayne et al., 2005), the additional degree of freedom in SMPC-ic allows for better transient closed-loop performance. While Tab. 1 does not include the $\ell_\xi$ component of the cost, SMPC-$\ell_\xi$ still preforms worse, since it cannot benefit freely from the additional degree of freedom as SMPC-ic.

Of course, this is not the purpose of using the additional $\ell_\xi$ cost, instead it causes the event $\xi \approx 0$ to occur more often, which can be observed in the example. However, the benefit of having the constraint conditioned on a more recent measurement, does not appear in the example as the constraint is anyway only active immediately after initialization.

## 7. CONCLUSION

We propose a stochastic MPC scheme using an optimization over the initial state for the predicted trajectory. For linear time-invariant systems with general additive stochastic disturbances, we have shown that standard guarantees can still be obtained. By using initial state of the prediction as an optimization variable, handling unbounded distribution becomes easily possible. Using a constraint tightening based on probabilistic reachable sets and assuming unimodal disturbances, we have shown closed-loop constraint satisfaction. Stability and cost decrease can be shown using standard methods enabling the adaption for many common cost functions. The simulation example highlights further benefits of the additional degree of freedom in the optimization such as improved transient behavior and performance. Future work will be focused on improving the performance and reducing conservatism through the use of growing tubes and larger constraint sets for the initial state in the prediction. Additionally, this could be applied to more general nonlinear system classes.